\documentclass{article}
\usepackage[a4paper,left=2.5cm,right=2.5cm,top=2.5cm,bottom=2cm, bindingoffset=0mm]{geometry}
\usepackage{amsfonts}
\usepackage{dsfont}
\usepackage{mathtools}
\usepackage{bm}
\usepackage{amsthm}
\usepackage{hyperref}

\usepackage{multirow}
\DeclareMathOperator{\Tr}{Tr}
\author{Andreas Mielke\footnote{mielke@tphys.uni-heidelberg.de}
  \\Institut f{\"u}r Theoretische Physik
  \\Ruprecht-Karls-Universit{\"a}t Heidelberg
  \\Philosophenweg 12
  \\D-69120 Heidelberg, Germany}
\title{Quantum Parrondo Games in Low-Dimensional Hilbert Spaces}
\date{\today}
\begin{document}
\maketitle

\begin{abstract}
We consider quantum variants of Parrondo games on low-dimensional
Hilbert spaces. The two games which form the Parrondo game
are implemented as quantum walks on a small cycle of length $M$.
The dimension of the Hilbert space is $2M$. We investigate a random
sequence of these two games which is realized by a quantum coin, so
that the total Hilbert space dimension is $4M$. We show that
in the quantum Parrondo game constructed in this way a systematic win
or loss occurs in the long time limit. Due to entaglement
and self-interference on the cycle, the game yields a rather
complex structure for the win or loss depending on the parameters.

\end{abstract}

\section{Introduction}
\label{sec:org17068b9}

Since their initial description by Harmer and Abbott \cite{Harmer_Abbott_1999},
\cite{Abbott_Harmer_1999} in 1999, classical Parrondo games attracted
a lot of attention and many different variants have beem proposed. In
all these variants, two simple fair (or losing but almost fair)
games are played in some regular or random sequence leading to
a systematic win. Parrondo invented these games as a simple, discrete
illustration of so called Bownian motors, see \cite{Reimann2001} for
a review. Other applications of Parrondo games are winning strategies
in social groups \cite{Cheong2020} or by bacteriophages \cite{Cheong2022}.

Quantum games have first been proposed by Meyer \cite{meyer1999quantumstrategies}
in 1999. 
Already three years later, Meyer and Blumer \cite{meyer-2002} proposed
quantum variants of Parrondo games based on lattice gas automata.
Later, Flitney \cite{flitney2012quantumParrondo} proposed a
different variant based on quantum walks. This approach is more
natural since the original Parrondo games can be viewed as
combinations of classical random walks. And their advantage
compared to the initial proposal
by Meyer and Blumer \cite{meyer-2002} is that they are
computationally less expensive. On the other hand, and this
holds for the different variants of capital dependent quantum Parrondo games
presented in the review \cite{Lai2020parrondoreview} by Lai and Cheong,
the Parrondo effect appears as a transient effect in these games
and vanishes in the infinite time limit. In contrast, Rajendran and Benjamin
\cite{Rajendran2018} show that if one uses two coin quantum walks,
one obtains a quantum Parrondo effect in the long time limit.
More recently, Panda {\emph et al.} \cite{Panda2022} show that
deterministic Parrondo sequences generate highly entangled states.
Recently, Walczak and Bauer  \cite{Walczak2021parrondosparadoxquantumwalks}
show that the quantum Parrondo effect occurs not only in random or
deterministic periodic sequences of games but also in deterministic
aperiodic sequences. 

A quantum game consists of an initial state and a series of unitary
transformations applied to that state.
The superposition of initial states, the quantum entanglement
of initial states, and quantum interference of the states induced
by the unitary transformations are the three important points which
make quantum games different from classical games. This holds for the
quantum Parrondo games as well.

In all the realization of quantum Parrondo games by quantum walks,
the games are realized using a quantum coin, \emph{i.e.} an
\(SU(2)\) transformation of a quantum bit, which is coupled to
a one-dimensional discrete walk on a line. The dimension of the Hilbert space
of this combination is infinite. On the other hand, the most
simple classical Parrondo game can be realized as a random walk
on a cycle with length 3. The classical game on this cycle
converges towards a stationary state which has a stationary, positive
current. Therefore the questions arises whether it is possible
to construct a quantum game using a quantum walk on a cycle
of finite length, \emph{e.g.} of length 3. The Hilbert space
of such a game is finite, depending on the implementation it
has dimension 6 or 12. This makes computations easy.
Further, Tregenna \emph{et al.} \cite{Tregenna2003controllingdiscrete}
point out that there is a fundamental
difference between quantum walks on the line and on the cycle.
For the latter, the interference of the quantum state with
itself makes a huge difference. This might make the investigation
of quantum Parrondo games on a cycle even more interesting.

The aim of this paper is to investigate quantum Parrondo games on a cycle.
The paper is structured as follows:
We first very briefly review classical Parrondo games on cycles.
We concentrate on capital dependent games and realizations where
the two different games are combined in random sequences.
We then explain how these games can be realized as quantum games
on cycles of length \(M\).
We keep this realization general, but we calculate explicit results
for cycles of length \(M=3\).
Generalizations to larger \(M\) are straight forward.
Since the quantum games do not converge to a
stationary state, we calculate the long time average of the current.
We show that quite complex patterns for the current depending on the
parameters of the game, esp. on the parameters of the quantum coin
occur. We show that, depending on the parameters, the maximal current
in the quantum system is larger than in the classical case and that
as a function of the parameters the current changes the sign.
Finally we give an outlook and propose some future research.

\section{Classical Parrondo games}
\label{sec:orgc1086d1}

The original, classical Parrondo game consists of two simple games,
game A and game B,
typically realized by flipping biased coins, which are played in some
regular or randomly alternating sequence. Since the quantum games
we investigate in this paper are realized by quantum walks, we use
the language of random walks for the classical games as well. In this section,
we very briefly introduce the original Parrondo game, mainly to fix
the notation and to compare the results for the quantum case later
with the classical results.

Game A
is a classical random walk on \(\mathbb{Z}\).
At time \(t=0\) the particle starts at \(x_0=0\).
At discrete time steps
\(t \in \mathbb{N}\), the particle moves
with probability \(1/2\) one step to the left or to the right.
The location of the particle after \(t\) time steps is \(x_t\).
The game is fair, the expectation value of \(x_t\) is 0.
The game is even a martingale since at every step it is fair.

Game B is a modified random walk. If \(x_t \bmod M =0\), the
particle moves with probability \(1/2-\epsilon_1\) to the right
and with probability \(1/2+\epsilon_1\) to the left.
If \(x_t \bmod M \neq 0\), the
particle moves with probability \(1/2+\epsilon_2\) to the right
and with probability \(1/2-\epsilon_2\) to the left. \(\epsilon_{1,2}\)
are chosen such that this game is fair as well.
Fair here means that in the long time limit no systematic win or loss
occurs. We will make this precise later.
The original choice by Harmer and Abbott \cite{Abbott_Harmer_1999}
was \(M=3\), \(\epsilon_1=2/5\), \(\epsilon_2=1/4\). 
This game is not a martingale since the single steps
are not fair. A fair game is only obtained in the stationary state.
One can prove that in the long-time limit the game converges
to the stationary state.

The classical Parrondo game (game C) is a random combination of game A and
game B. One plays game A with probability \(p_{\mathrm A}\) and game B
with probability \(p_{\mathrm B}=1-p_{\mathrm A}\). Doing so, one
obtains a game which is no longer fair. The random walker moves
more often in one direction than in the other. This effect
is called Parrondo's paradox.

More formally, one can treat the system as a discrete time Markov process
\cite{Markov_Chains}
on a reduced state space \(\mathbb{Z}/M\mathbb{Z}\). Let \(p_{t,x}\),
\(x \in \mathbb{Z}/M\mathbb{Z}\) be the probability for the particle
to be in state \(x\) at time \(t\).
For the games A, B, or C the
Markov process is described by a stochastic matrix
\(L_{\mathrm A, \mathrm B, \mathrm C}\)
and 

\begin{equation}
p_{t+1}=L_{\mathrm A, \mathrm B, \mathrm C}p_t
\end{equation}

where \(p_t=(p_{t,x})_{x \in \mathbb{Z}/M\mathbb{Z}}\). We have
\begin{equation}\label{transition matrix}
  L_{\mathrm B} = \left(\begin{array}{ccccc}
                     0        & 1/2-\epsilon_2 &  &  & 1/2+\epsilon_2\\
                     1/2-\epsilon_1  & 0         &\ddots&&\\
                              & 1/2+\epsilon_2    & \ddots & 1/2-\epsilon_2&\\
                              &               &\ddots&0& 1/2-\epsilon_2\\
                     1/2+\epsilon_1      &    &  & 1/2+\epsilon_2 & 0
       
                   \end{array}\right)\, ,
\end{equation}
the same for \(L_{\mathrm A}\) but with \(\epsilon_{1,2}=0\), and
\(L_{\mathrm C}=p_{\mathrm A}L_{\mathrm A}+p_{\mathrm B}L_{\mathrm B}\).

If \(\rho\) is the stationary solution
(right eigenvector of \(L\) with eigenvalue 1),
then let \(P=\mathrm{diag}\,\rho\).
The stationary current then is
\begin{equation} \label{jclassical}
j=LP-PL^t \, .
\end{equation}
\(j_{x,y}\) is the probability for the particle to sit on \(y\) and move to \(x\)
minus the probability to sit on \(x\) and move from \(x\) to \(y\).
\(j_{x,y}\) is therefore the net flow from \(y\) to \(x\).
Detailled balance means \(j=0\), this is the case for game A.
For a fair game it is sufficient that
\begin{equation}
\sum_y j_{x,y}=0\, .
\end{equation}
The parameters in game B are tuned such that it is fair.
For game B with \(M=3\) it is easy
to show that the game is fair if
\(\epsilon_1=2\epsilon_2/(4\epsilon_2^2+1)\).
The values 
\(\epsilon_1=2/5\), \(\epsilon_2=1/4\)
chosen by Harmer and Abbott \cite{Abbott_Harmer_1999}
clearly fulfill this condition.

\begin{figure}[ht]
\includegraphics[width=0.75\textwidth]{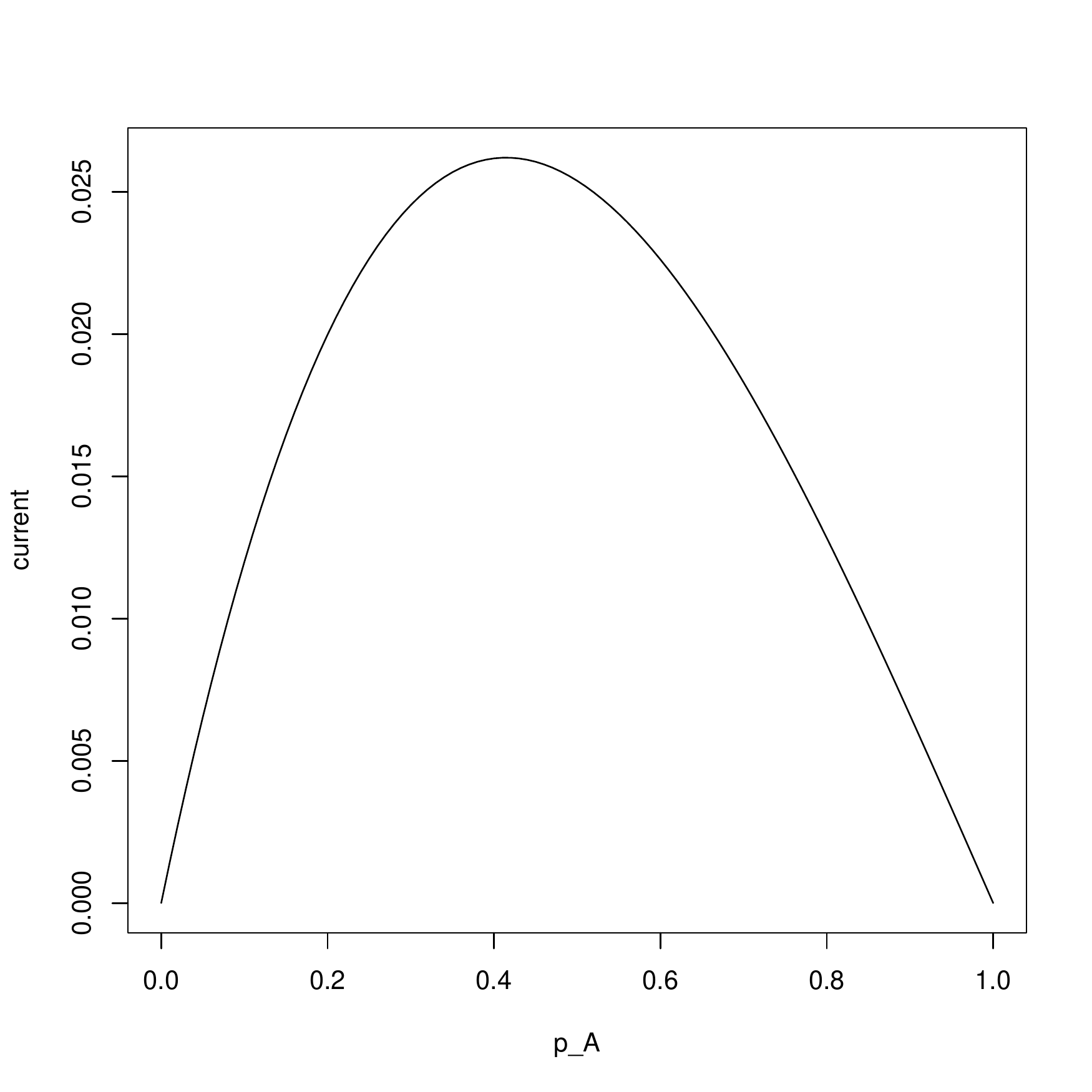}

\caption{\label{fig:StationaryCurrent}The stationary current for
 $M=3$, $\epsilon_1=2/5$, $\epsilon_2=1/4$.}
\end{figure}

Calculating the current for the combination, game C, is easy.
One just needs to calculate the eigenvector of \(L_{\mathrm C}\)
and use (\ref{jclassical}).
Fig.  \ref{fig:StationaryCurrent} shows the result for the
current as a function of \(p_{\mathrm A}\) for the original parameter
set  \(M=3\), \(\epsilon_1=2/5\), \(\epsilon_2=1/4\).

Since the two games A and B are fair, the current vanishes for \(p_{\mathrm A}=0\)
and \(p_{\mathrm A}=1\). A maximum occurs near \(p_{\mathrm A}=0.4\).

This result is well known. One can modify \(M\), one can introduce different
transition probabilities for the switch from game A to game B and back,
one can tune \(\epsilon_{1,2}\), etc. We do not go further here because our
interest is to investigate a quantum version of this game.

\section{Quantum Parrondo games}
\label{sec:orgde0da22}

As an implementation for the game we use a quantum walk.
It consists of a single quantum coin realized by a quantum bit and a
chain (or cycle). Two unitary transformations are performed in one time step.
The first is tossing the coin, the second is moving the walker along the chain
depending on the result of the coin. If the walker walks on a cycle of length
\(M\), we need therefore a Hilbert space of dimension \(2M\).

A quantum coin can be realized by elements of the
(two-dimensional representation of the)
Lie group \(SU(2)\). 
We choose an arbitrary \(SU(2)\) matrix. We use the parametrization
\begin{equation} \label{eq:paramV}
V(q,\phi,\theta)=\begin{pmatrix}
\sqrt{q}\exp(i\phi) & \sqrt{1-q}\exp(i\theta) \\
-\sqrt{1-q}\exp(-i\theta) & \sqrt{q}\exp(-i\phi)
\end{pmatrix}
\end{equation}
with \(0\leq q \leq 1\), \(0\leq \phi \leq 2\pi\), \(0\leq \theta \leq 2\pi\).

Let \(J\) be a unitary transformation for a move
of the walker to the right, \(J^\dagger\)
for a move to the left, and let \(I\) be the identity.
Let now \(V\in SU(2)\) be the unitary coin tossing transformation.
Let \(P_\pm\) be the projectors onto the two states of the coin.
Then
\begin{equation}
U_\mathrm{A}=(J \otimes P_+ + J^\dagger \otimes P_-) (I \otimes V)
\end{equation}
represents the motion of the quantum walker. This would be the realization
of game A of a Parrondo game provided the parameters are chosen such
that the game is fair, esp. we need \(q=1/2\) as in the classical case.

The question under which conditions \(U_\mathrm{A}\)
is unbiased (or fair) has been investigated by
Tregenna \emph{et al.} \cite{Tregenna2003controllingdiscrete}.
We reproduce their result here in a slightly modified way
because we need to generalize it later when we deal with game B.

\(\langle P_\pm\rangle\) are the expectation values for the two
states of the coin. At a given time step \(t\), the system is
in a given state described by a density operator \(\rho_t\).
\begin{equation} 
\langle (P_+ - P_-)\rangle_t = \Tr (P_+ - P_-)\rho_t
\end{equation}
yields the net flow, namely the probability to move to the right
minus the probability to move to the 
left. We calculate this quantity for each time
step and take the average over all time steps, \emph{i.e.}
\begin{equation} \label{jquantum}
j:=\langle (P_+ - P_-)\rangle =
\lim_{T\rightarrow\infty}\frac{1}{T} \sum_{t=1}^T \Tr (P_+ - P_-)\rho_t\, .
\end{equation}
In the classical case, the long time average is identical to the
stationary current calculated there because in the classical
case the stochastic process converges to the stationary solution.
For the quantum walk there is generically no convergence
to a stationary state, therefore we take the long time average.

For a general game which is defined by a unitary transformation \(U\)
(take \(U_\mathrm{A}\) as an example), 
we have \(\rho_t = U^t \rho_0 (U^\dagger)^t\).
Therefore we can calculate the trace in (\ref{jquantum}) in the eigenbasis of \(U\).
Let \(L\) be the set of eigenvalues of \(U\)
and let for \(\exp(i\lambda)\in L\) be \(P_\lambda\) be the projector
onto the eigenspace of the eigenvalue \(\exp(i\lambda)\).
Then we obtain
\begin{equation} \label{eq:j}
j = \sum_{\exp(i\lambda)\in L}
\Tr((P_+ - P_-)P_\lambda\rho_{0}P_\lambda)\, .
\end{equation}
An unbiased (or fair) game is a game where \(j=0\).
The game is biased or unbiased with respect to a given
\(\rho_0\). There is no parameter set for which the game is unbiased
for arbitrary \(\rho_0\).

We are mainly interested in the case where \(\rho_0\) is a pure state.
But it is clearly possible to do the calculation for mixed states.
If \(\rho_0\) is proportional to the identity, we have a fully mixed state.
For this case, \(\rho_t=\rho_0\) and \(j\) vanishes. This holds for arbitrary \(U\).

The eigenbasis of \(U_\mathrm{A}\) is easily calculated.
Let \(u\) be an eigenstate of \(J\)
to the eigenvalue \(\exp(i\xi)\). Since \(J\) and \(J^\dagger\) commute,
we can write the eigenstates of \(U_\mathrm{A}\) in the form
\begin{equation}
\begin{pmatrix}
z_+ u\\
z_- u
\end{pmatrix}
\end{equation}
with two complex numbers \(z_\pm\). Those are determined by the eigenvalue
equation
\begin{equation}
\begin{pmatrix}
\sqrt{q}\exp(i\phi+i\xi) & \sqrt{1-q}\exp(i\theta+i\xi) \\
-\sqrt{1-q}\exp(-i\theta-i\xi) & \sqrt{q}\exp(-i\phi-i\xi)
\end{pmatrix}
\begin{pmatrix}
z_+\\
z_- 
\end{pmatrix}
=\exp(i\lambda)
\begin{pmatrix}
z_+\\
z_- 
\end{pmatrix}\, .
\end{equation}
For each eigenvalue \(\exp(i\xi)\) of \(J\) we get two eigenvalues \(\exp(i\lambda)\)
of \(U_\mathrm{A}\). 
From the eigenvalue equation we obtain two equations for \(z_\pm\) and
those can be reduced to
\begin{equation}
\cos(\lambda) = \sqrt{q} \cos(\phi + \xi)
\end{equation}
and 
\begin{equation}
\sqrt{1-q}\exp(-i\theta)z_+ = (\sqrt{q}\exp(-i\phi) - \exp(i\lambda+i\xi))z_-\, .
\end{equation}

So far we have not been specific about the set the quantum walker
is defined on. As stated and motivated in the introduction, we use a cycle.
Tregenna \emph{et al.} \cite{Tregenna2003controllingdiscrete} point out that
there is a fundamental
difference between quantum walks on the line and on the cycle.
For the latter, the interference of the quantum state with
itself makes a huge difference. The interference is one motivation
to study the games on a cycle, but more importantly, it makes
the calculations cheap since the Hilbert space dimension is small
for small cycles. For a cycle of length \(M\) the Hilbert space dimension
is \(2M\) and the smallest possible value of \(M\) is \(M=3\).

For a cycle of length \(M\) one has \(J^M=1\) and therefore
the eigenvalues of \(J\) are \(\exp(i\xi_l)\) with
\(\xi_l=2\pi l/M\), \(|l|\leq M/2\). For a pair of two \(\xi\), \(\xi'\)
for which \(\xi\) and \(\phi+\xi'\) differ only by a sign,
we obtain the same two values for \(\lambda\)
for each. Thus there may be degeneracies in the spectrum of \(U_\mathrm{A}\).

For the concrete calculations, we use
the smallest possible cycle of length \(M=3\). 
We choose
\begin{equation}  \label{eq:InitialConditionWalk}
\rho_0=\rho_{0w} \otimes \rho_{0c}
\end{equation}
with
\begin{equation}\label{eq:InitialConditionCoin}
(\rho_{0w})_{x,y \in \mathbb{Z}/M\mathbb{Z}} =\delta_{x,y}\delta_{x,0}
\end{equation}
and
\begin{equation}
\rho_{0c}=\frac{1}{2}
\begin{pmatrix}
1 & \pm 1 \\
\pm 1 & 1
\end{pmatrix}\, .
\end{equation}
In \(\rho_{0c}\) we usually choose the upper sign, but the results
we present here are independent of the sign. Note that \(\rho_{0c}\)
is the projection onto the state \((1, \pm 1)^\dagger/\sqrt{2}\),
\(\rho_0\) therefore is a density matrix of a pure state.
Generalizations to mixed states are straight forward.

\begin{figure}[ht]
\includegraphics[width=0.75\textwidth]{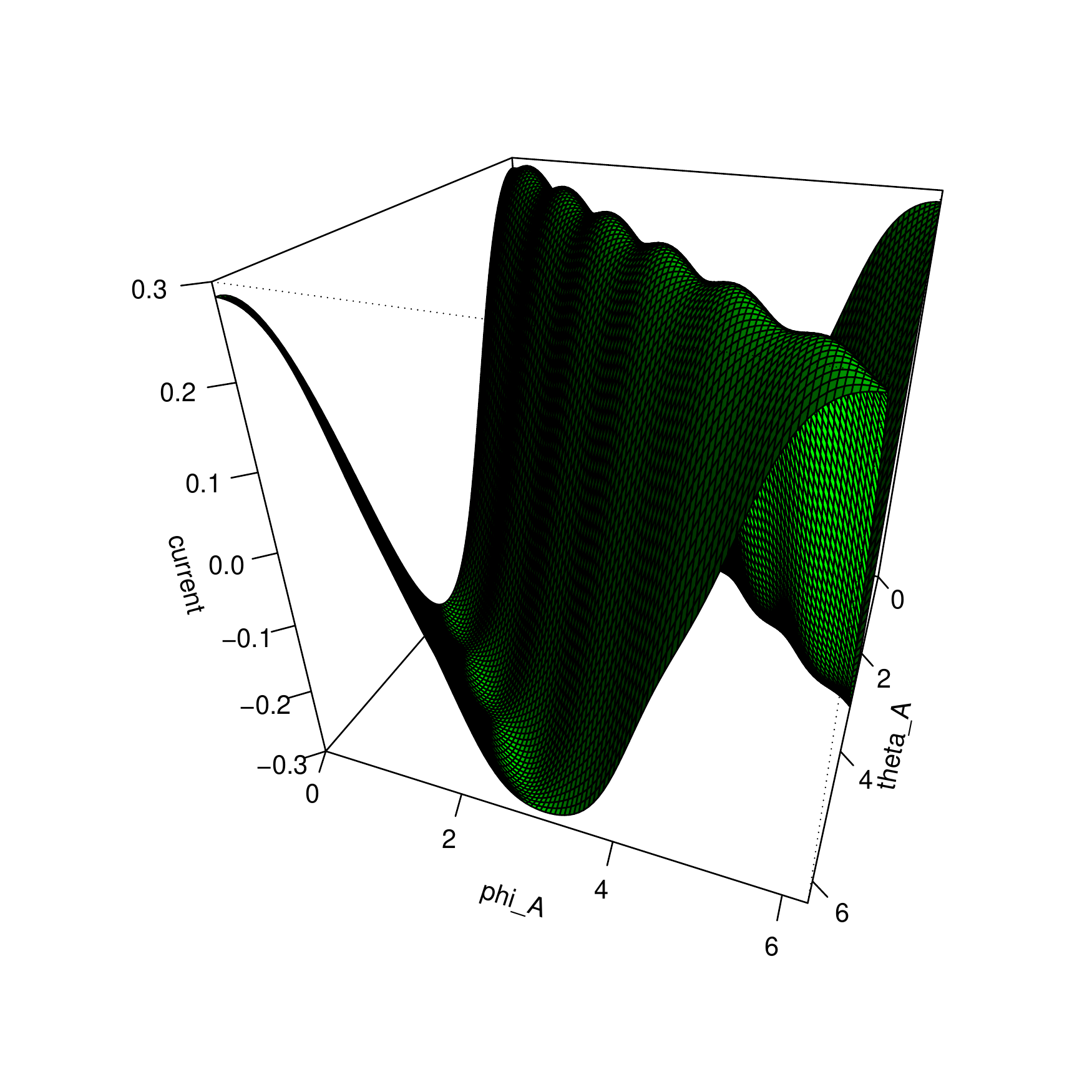}

\caption{\label{fig:CurrentQA}The time averaged current for
 the quantum game A, $M=3$.}
\end{figure}

In the \(SU(2)\) matrix \(U_{\mathrm A}\) which represents the quantum coin,
the probability is fixed to \(q_{\mathrm A}=1/2\) for which the classical game A
is fair. The quantum game has two free parameters, the  two phases
\(\phi_{\mathrm A}\) and \(\theta_{\mathrm A}\) in the representation
of the matrix \(V\) in (\ref{eq:paramV}).
The calculation of the current is easy, we just need to calculate
the eigenvalues of the unitary matrix \(U_{\mathrm A}\) and the corresponding
projectors onto the eigenspaces and apply (\ref{jquantum}).
The result for the current of game A as a function of the two phases
\(\phi_{\mathrm A}\) and \(\theta_{\mathrm A}\) of \(U_{\mathrm A}\)
is shown in Fig. \ref{fig:CurrentQA}. The
current vanishes for \(|\theta_{\mathrm A}-\phi_{\mathrm A}|=\pi/2 \bmod \pi\).
One possible choice to obtain a fair game is \(\phi_{\mathrm A}=0\)
and \(\theta_{\mathrm A}=\pi/2\). 
This has as well been shown in \cite{Tregenna2003controllingdiscrete}.

For game B of a Parrondo game we need two different unitary coin tossing transformations
\(V_{1,2}\in SU(2)\). Which one is used depends on the site of the walker.
Let \(P_{1,2}\) be the projector onto the two subsets of walker sites
for which those two coin transformations are used, \(P_1 + P_2 = I\). Then the unitary
transformation for game B is

\begin{equation}
U_\mathrm{B}=(J \otimes P_+  + J^\dagger \otimes P_- )(P_1 \otimes V_1 + P_2 \otimes V_2)
\, .
\end{equation}
For \(U_B\) we choose two different probabilities \(q_1=1/2-\epsilon_1\)
and \(q_2=1/2+\epsilon_2\) for the
two matrices \(V_1\) and \(V_2\) as in the original
classical Parrondo game.
We choose \(M=3\) and \(\rho_0\) as above for game A. 
Further we use the same phases \(\phi_{\mathrm B}\) and \(\theta_{\mathrm B}\)
in \(V_1\) and \(V_2\).

\begin{figure}[ht]
\includegraphics[width=0.75\textwidth]{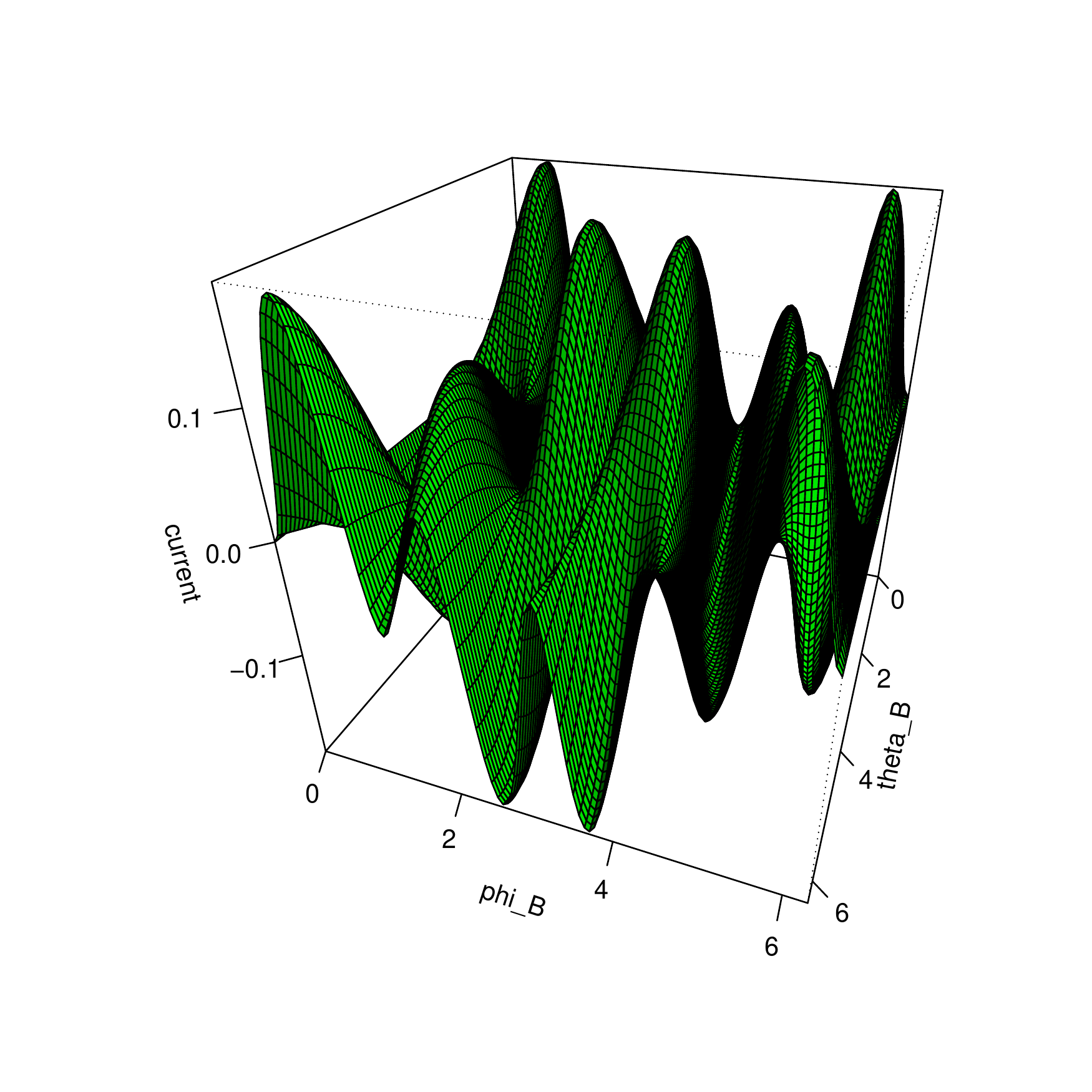}

\caption{\label{fig:CurrentQB}The time averaged current for
 the quantum game B, $M=3$.}
\end{figure}
The result for the current of game B as a function of the two phases
\(\phi_{\mathrm B}\) and \(\theta_{\mathrm B}\) is shown in Fig. \ref{fig:CurrentQB}. The
current vanishes for a larger set of parameters than for game A.
One possible choice to obtain a fair game is \(\phi_{\mathrm B}=0\) and
\(\theta_{\mathrm B}=\pi/2\),
another possible choice is \(\phi_{\mathrm B}=\pi/2\) and \(\theta_{\mathrm B}=0\).
For the following calculations we take the latter choice.

\begin{figure}[ht]
\includegraphics[width=0.75\textwidth]{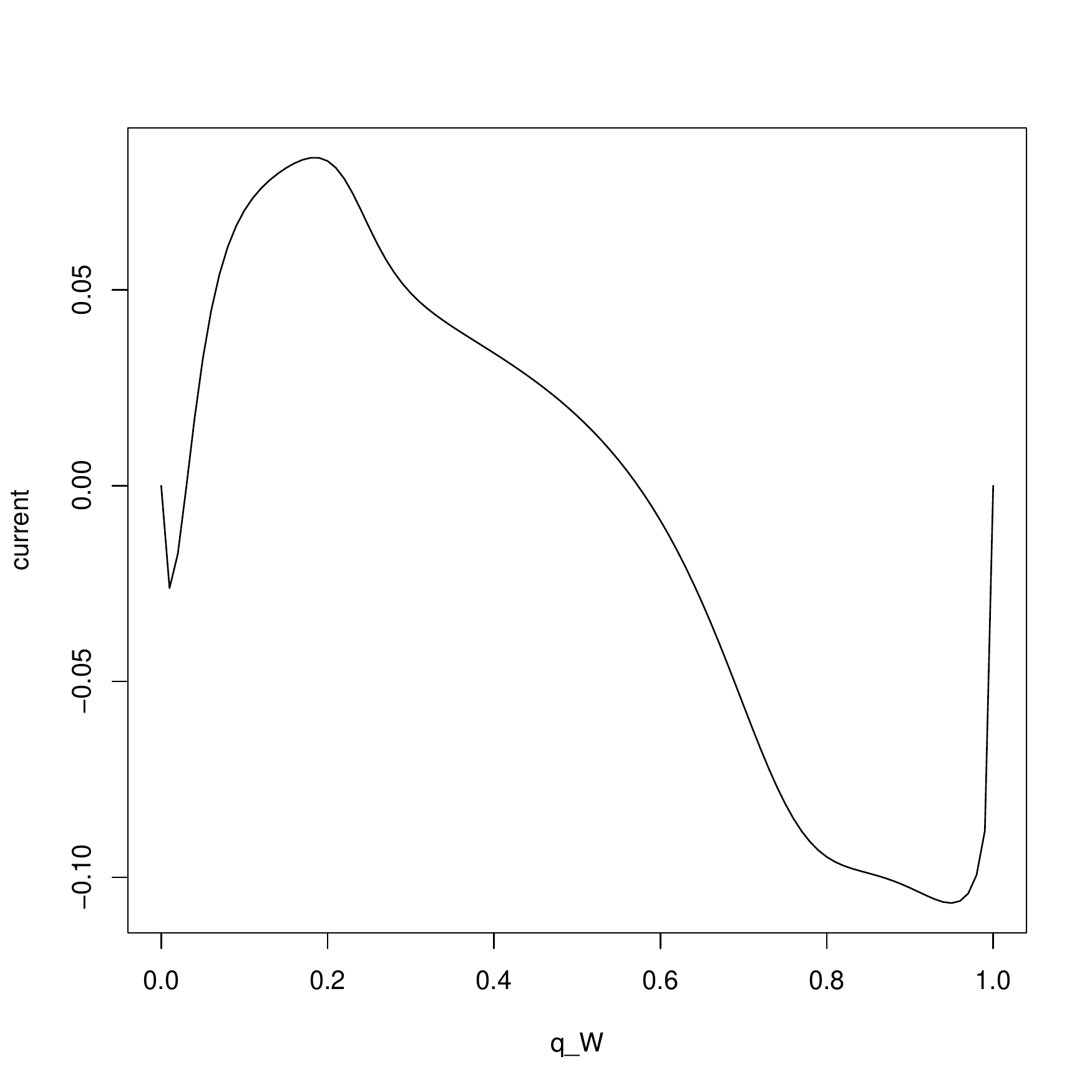}

\caption{\label{fig:parrondoQuantumJg-q}The time averaged current for
 the combination of the two unbiased games A and B with $\theta_{\mathrm A}=\pi/2$,
 $\phi_{\mathrm A}=0$, $\theta_{\mathrm B}=0$, $\phi_{\mathrm B}=\pi/2$.
 The current is plotted as a function of the probability $q_W$ in the unitary
 transformation $W$ for the choice of game A, $\theta_W=\pi/2$, $\phi_W=0$.
}
\end{figure}

To obtain a Parrondo game, game A and B are played either in some regular,
alternating fashion
or randomly. We concentrate on the latter case, for which we need a second qbit.
On the cycle, the Hilbert space dimension is then \(4M\).
Let \(W\in SU(2)\) represent the unitary transformation for the second
qbit and let \(Q_{\pm}\) be the two projectors onto the two basis states
of the second qbit. Then the entire unitary transformation for the
quantum Parrondo game is
\begin{equation}\label{eq:U}
U=(U_A \otimes Q_+  + U_B \otimes Q_- )(I \otimes W)\, .
\end{equation}
This corresponds to the classical game C.
The construction is easy, \(U\) consists of two unitary transformations.
The first is the coin tossing represented by \(W\), the second is the
unitary transformation of game A or game B depending on the state of the coin.
As before the current of the combined game is easily calculated
by calculating the eigenvalues and the corresponding projectors
of the unitary matrix \(U\).
We obtain generically a bias, \emph{i.e.} \(j\neq 0\). The value and the sign of the
current depend on the parameters \(q_W\), \(\phi_W\) and \(\theta_W\)
in \(W\) in the representation (\ref{eq:paramV})
and on the initial conditions. We choose the same initial
condition as before in games A and B and we fix the parameters of the
quantum coines in game A and B as described above, so that these two games are
fair, meaning that the long time average of the current is zero.
For the initial condition of the second quantum coin we take
\begin{equation}
\rho_{0W}=\frac{1}{2}
\begin{pmatrix}
1 & \pm 1 \\
\pm 1 & 1
\end{pmatrix}
\end{equation}
Again, we choose the upper sign. For the lower sign
we obtain similar results. 
Then, the combined game depends on the three parameters  of \(W\).
We first take \(\theta_W=\pi/2\), \(\phi_W=0\) and calculate the current as a
function of \(q_W\). The result is plotted in Fig. \ref{fig:parrondoQuantumJg-q}.
The current is negative for very small \(q\), then changes the sign and becomes
positive with a maximum at \(q\approx 0.2\). It then decreases, becomes negative
with a maximum for \(q\approx 0.9\). The current vanishes for \(q=0\) and \(q=1\).
This must be the case since then only game A or only game B is played
and both have a vanishing current.

\begin{figure}[ht]
\includegraphics[width=0.75\textwidth]{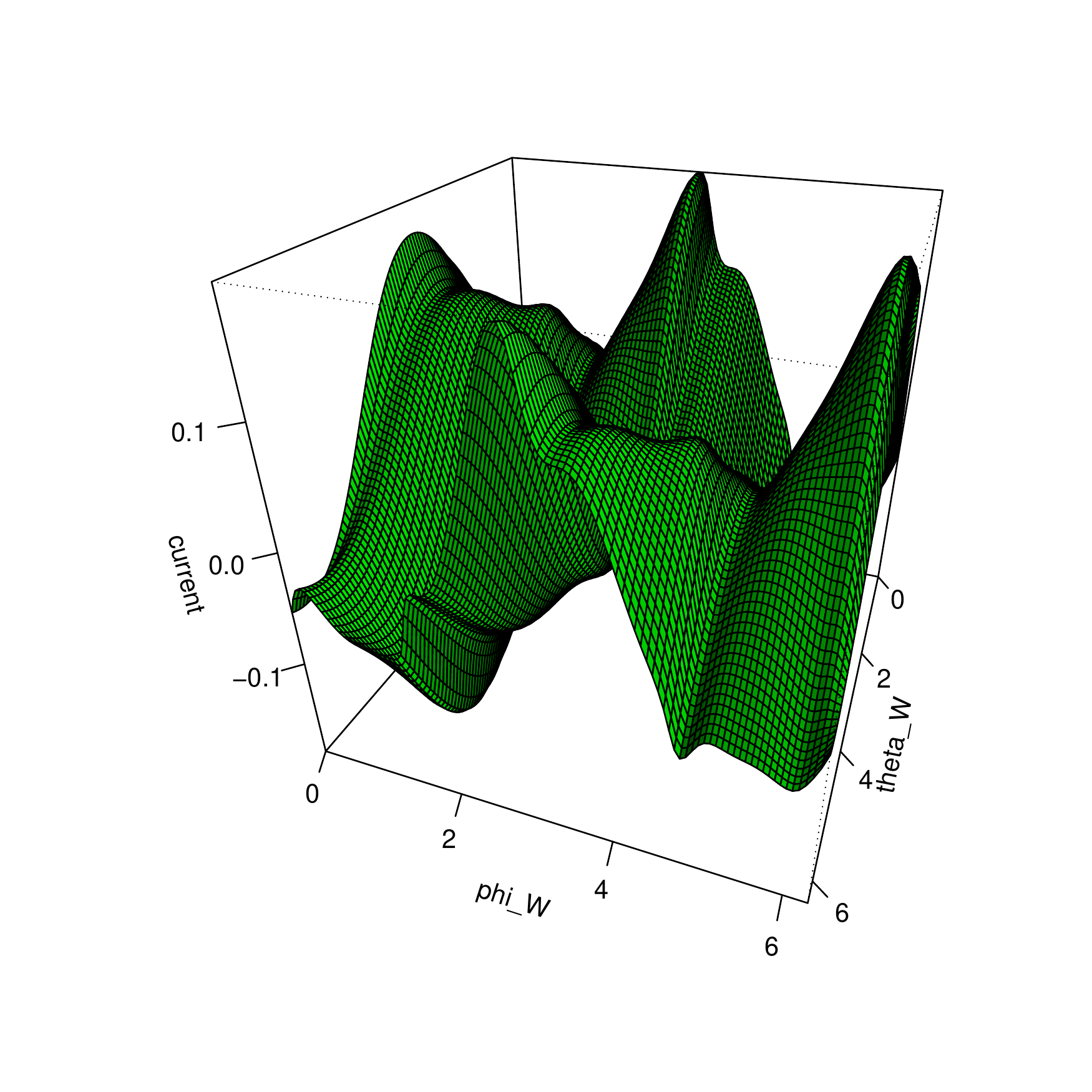}

\caption{\label{fig:quantumCJg-theta-phi-q0.2}The time averaged current for
 the combination of the two unbiased games A and B with $\theta_{\mathrm A}=\pi/2$,
 $\phi_{\mathrm A}=0$, $\theta_{\mathrm B}=0$, $\phi_{\mathrm B}=\pi/2$.
 The current is plotted as a function of the two phases $\theta_W$ and $\phi_W$
 and a fixed probability $q_W=0.2$ in the unitary
 transformation $W$ for the choice of game A.
}
\end{figure}

\begin{figure}[ht]
\includegraphics[width=0.75\textwidth]{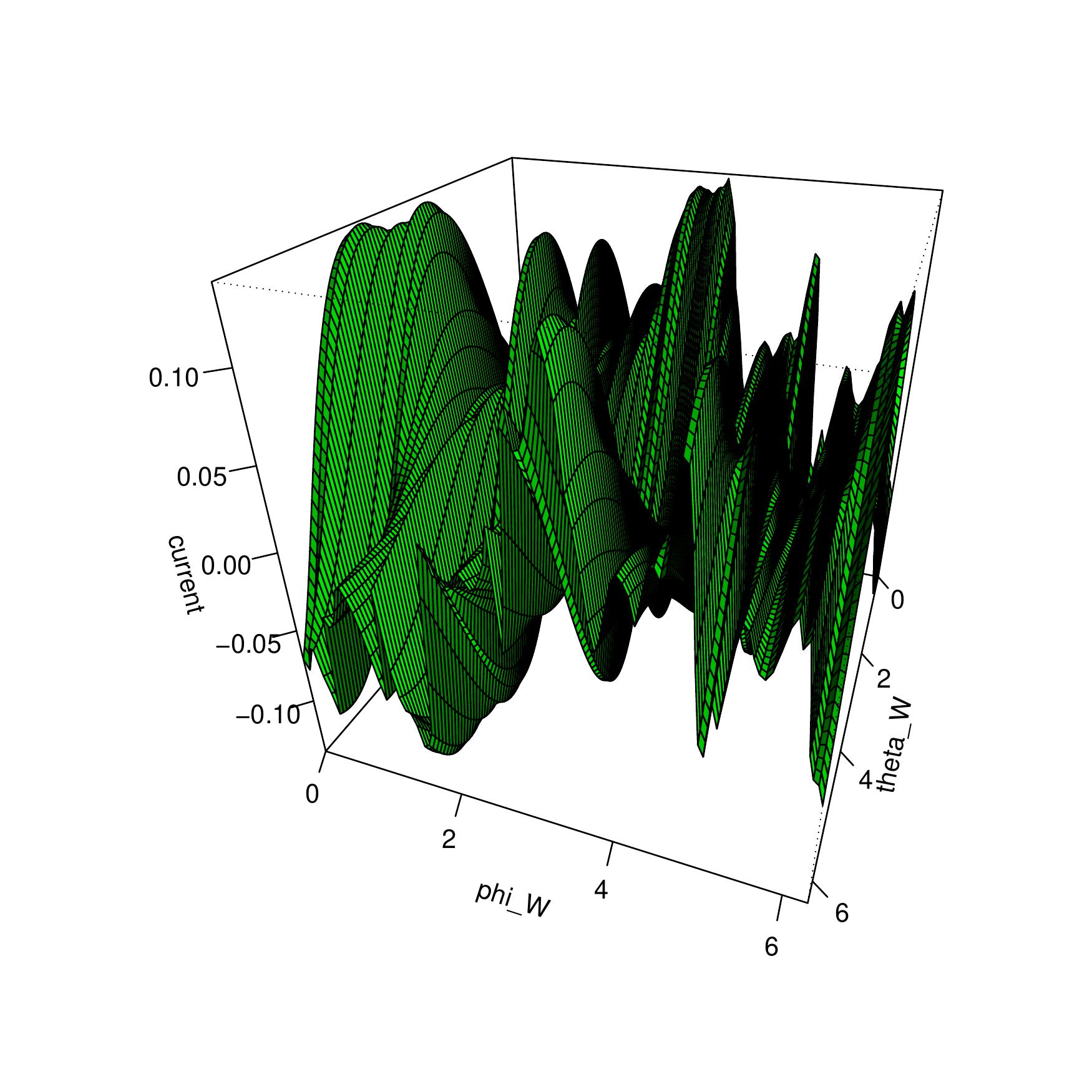}

\caption{\label{fig:quantumCJg-theta-phi-q0.8}The time averaged current for
 the combination of the two unbiased games A and B with $\theta_{\mathrm A}=\pi/2$,
 $\phi_{\mathrm A}=0$, $\theta_{\mathrm B}=0$, $\phi_{\mathrm B}=\pi/2$.
 The current is plotted as a function of the two angles $\theta$ and $\phi$
 and a fixed probability $q_W=0.8$ in the unitary
 transformation $W$ for the choice of game A.
}
\end{figure}

We now fix \(q_W\) and calculate the current as a function of the
phases \(\theta_W\) and \(\phi_W\). We show the result for \(q_W=0.2\)
in Fig. \ref{fig:quantumCJg-theta-phi-q0.2} and the result for \(q_W=0.8\)
in Fig. \ref{fig:quantumCJg-theta-phi-q0.8}. In both cases
we see a very rich structure of the current as a function of
the two phases with many sign changes, minima, and maxima. The
maxima of the absolute value of the current lie above 0.1,
a factor 5 higher than in the classical case.

The rich structure we observe here is a lot more complicated than 
the structure of the single games A and B depicted in Figs. \ref{fig:CurrentQA},
\ref{fig:CurrentQB}. What actually happens is the following: We have a
quantum walk on a cycle which interferes with itself during its motion.
Actually, this quantum walk is an \(M\) state quantum system with \(M=3\)
in our concrete calculations.
It is coupled to two quantum coins, each of dimension 2.
These three quantum systems evolve in discrete time steps
with a unitary transformation (\ref{eq:U}). The transformation
produces a sequence of entangled states which clearly depend
on the parameters of the transformation. The entanglement and the interference
between of these states with themselves produce the rich structure
of the long-time average of the current.
This is the main result of the paper: It is possible to construct
quantum Parrondo games on small Hilbert spaces, they generically show
a finite current which may be positive or negative and which is larger
than the classical current, and the current as a function of the parameters
has a rich structure.

This is not always the case. The effect occurs for our choice of the phases of games
A and B, \(\theta_{\mathrm A}=\pi/2\),
\(\phi_{\mathrm A}=0\), \(\theta_{\mathrm B}=0\), \(\phi_{\mathrm B}=\pi/2\)
and for many other choices for which the games A and B are fair.
But there are exceptions. 
If instead we choose \(\theta_{\mathrm A}=\theta_{\mathrm B}=\pi/2\),
\(\phi_{\mathrm A}=\phi_{\mathrm B}=0\),
the current vanishes independent of the choice of \(W\).

\section{Discussion and outlook}
\label{sec:orgc4ab4ad}

In the classical case of game B, some fine tuning is needed to obtain a fair
game. The classical games are discrete time Markov processes, for a
mathematical introduction see \cite{Markov_Chains}. Whereas game A is fair because
it is a martingale (it is balanced at every single step), game B is not
a martingale and the fairness, a vanishing current in the long-time limit,
can only be assured by the fine tuning of the two probabilities involved.
Combining the classical games A and B randomly, without fine tuning, leads
to a current. Given the fact that already for the fairness of game B fine
tuning is necessary, the occurence of a current for the random combination
of games A and B is eventually less paradoxical than the term "Parrondo's paradox"
suggests.

Quantum walks, in contrast to classical random walks, are much more complicated
and do not even converge to a stationary state. Therefore, more fine tuning is needed,
even for the quantum analogue of game A to obtain a fair game, a game for which
the long time average of the current vanishes. A similar fine tuning is needed
for game B. That the random combination of the two games is not fair, is therefore
something which might have been expected. From this point of view, the inital
statement of Flitney \cite{flitney2012quantumParrondo} that the Parrondo effect
occurs only as a transient effect and vanishes in the long time limit, may look
unexpected. But Flitney simply combined two simple quantum walks of
the type of game A, not the quantum analogues to the classical game B. This may explain
his result.

An important point to understand the results for the quantum games is the
well known fact that fairness of the quantum games is not independent
of the initial condition. We choose the initial condition
(\ref{eq:InitialConditionCoin}) for the coin which is a pure entangled
combination of the two states of the quantum coin. If one chooses
as initial condition the normalized unit matrix, \emph{i.e.}
\begin{equation}
\rho_{0c}=\frac{1}{2}
\begin{pmatrix}
1 & 0 \\
0 & 1
\end{pmatrix}
\end{equation}
which is not a pure state
but the mixture of the two orthogonal pure states of the coin with equal weight,
the current is always zero because any unitary transformation \(V\) in (\ref{eq:paramV})
leaves this state invariant. In this case, there is no entanglement
and no interference. In other words: If we take the average of the current
for all possible different initial conditions, this should be zero.
If we take the average over a subset of states, namely the states
\(P_\lambda\rho_0P_\lambda\) in (\ref{eq:j}) we may generically get non-zero
values with different signs depending on the parameters which determine \(P_\lambda\),
\emph{i.e.} the parameters of the unitary transformation that defines the game.

The interesting result is that the depencence of the current has an
extremely rich structure as shown in Figs. \ref{fig:parrondoQuantumJg-q},
\ref{fig:quantumCJg-theta-phi-q0.2}, and \ref{fig:quantumCJg-theta-phi-q0.8}.
A second, equally important result is that, depending on the parameters,
the current may by much larger than in the classical case.

The rich structure we observe here certainly provokes a series of questions:
Which conditions are needed for \(U\) and the initial conditions to obtain
the rich structure we observe and is it possible to characterize this
structure in some way? Is it possible to prove that for certain conditions
the current vanishes? What happens if instead of taking a random quantum choice
of games A and B we play games A and B in a periodic sequence? What happens for
larger values of \(M\) and especially in the limit \(M\rightarrow\infty\)?
We investigate the long time behaviour of the current, what about
the transient behaviour, is it different? The early result by
Flitney \cite{flitney2012quantumParrondo} seem to suggest that.
These questions show that there is room for further research.

Another direction of research would be to investigate multi-player
versions of the quantum games presented here. For multi-player
versions of classical Parrondo games, interesting effects have
been observed, see \emph{e.g.} \cite{Breuer2023}. For the quantum
version, due to the rich struture we obtain, each of the players
has a huge flexibility to develop an optimal strategy.

\end{document}